\documentstyle[twoside,fleqn,espcrc2,epsfig]{article}

\title{Exact chiral symmetry, topological charge and related topics}

\author{Ferenc Niedermayer\address{Paul Scherrer Institute, Villigen,
        Switzerland}%
        \thanks{Present address: Institute for Theoretical Physics,
                University of Bern,
                on leave from E\"otv\"os University, Budapest}}

\begin{document}

\begin{abstract}
It has been shown recently that Dirac operators satisfying
the Ginsparg-Wilson relation provide a solution of the chirality
problem in QCD at finite lattice spacing. We discuss different
ways to construct these operators and their properties.
The possibility to define lattice chiral gauge theories is briefly
discussed as well.
\end{abstract}

\maketitle

\section{Introduction}
Chiral symmetry plays an essential role in particle physics. 
In QCD it is a global symmetry which forbids an additive quark 
mass renormalization and whose spontaneous breakdown provides us 
with (nearly) Goldstone bosons with their specific interactions.

On the other hand, in electroweak interactions chiral symmetry plays 
an even more fundamental role: it is a local gauged symmetry. 
Its presence is necessary to provide a renormalizable theory.

Accordingly, when trying to regularize these theories
on the lattice (which is the only non-perturbative regularization
known for gauge theories) one meets two basic problems.
As it is (or rather {\em was}) well known one cannot realize chiral
symmetry exactly on the lattice without sacrificing some even more
important principles, like locality or absence of extra physical
particles. This is the content of the famous
no-go theorem by Nielsen and Ninomiya \cite{NN}.
In Wilson's formulation of lattice QCD chiral symmetry is explicitly 
broken by an irrelevant term which does not affect
the continuum limit. It causes, however,  O$(a)$ lattice
artifacts and other unwanted effects at finite lattice spacing.

For the electroweak theory, on the other side, the problem of chiral
symmetry on the lattice is a principal one: is it possible
to define a chiral gauge theory non-perturbatively?
This second problem is much more difficult, and assumes
that the first one -- a solution for global chirality -- has been
successfully solved.

Several ways to solve these problems have been proposed. These
were mostly covered on previous lattice conferences.
Here I mention only the domain wall fermions \cite{DWF,Blum98}
and the related overlap formalism \cite{OLF,Neub98}.
On can state that these approaches provide a solution 
to the problem of global chirality. 
The numerical results using domain wall fermions \cite{Blum98}
are encouraging.
In the overlap formalism also important steps were made towards 
the solution of the gauged chiral symmetry 
(for references see Ref.~\cite{Neub98}).

It was unclear, however, to what extent are the theoretical constructions
used in these approaches necessary, or what is the basic ingredient
behind the chiral properties observed.
As it became clear recently, the basic relation which has to be satisfied
to realize chiral symmetry at finite lattice spacing, is the 
Ginsparg-Wilson (GW) relation \cite{GW}
suggested a long time ago!

In this talk I shall concentrate on the role of the GW relation
and its consequences. Some of the results were obtained earlier
in the approaches mentioned above, but even these are more transparent
within the new framework.

\subsection{The Nielsen-Ninomiya no-go theorem}

Consider the free fermionic action
\begin{equation}
  \label{SF}
  S_{\rm F}=a^4 \sum_{x,y} \bar{\psi}(x) D(x-y) \psi(y) \,,
\end{equation}
describing a massless fermion on the lattice. (Colour, Dirac and
flavour indices are suppressed.) 

The desirable properties of the Dirac operator $D$ should be
\begin{itemize}
\item[(a)] $D(x)$ is local (bounded by $C {\rm e}^{-\gamma |x|}$)
\item[(b)] $\tilde{D}(p)=i\gamma_\mu p_\mu +O(ap^2)$ for $p\ll \pi/a$
\item[(c)] $\tilde{D}(p)$ is invertible for $p\ne 0$ (no massless doublers) 
\item[(d)] $ \gamma_5 D + D \gamma_5 =0$
\end{itemize}
Here $\tilde{D}(p)$ is the Fourier transform of $D(x)$.

The theorem by Nielsen and Ninomiya \cite{NN} states that properties
(a)-(d) cannot hold simultaneously.

Following Wilson, to avoid the massless doublers present in the naive
discretization of the Dirac operator, one introduces 
a chiral symmetry breaking, irrelevant operator.
The corresponding Wilson-Dirac operator is given by
\begin{equation}
  \label{Dw}
  D_{\rm w}= \frac{1}{2} \gamma_\mu ( \nabla_\mu +\nabla^*_\mu)
  - \frac{1}{2} a \nabla^*_\mu \nabla_\mu \,,
\end{equation}
where $\nabla_\mu$ ($\nabla^*_\mu$) is the forward (backward) lattice
derivative.

Due to the explicit breaking of chiral invariance, however, some extra
problems appear:

(1) Chiral symmetry is broken at finite lattice spacing by 
O($a$) lattice artifacts and it is recovered only in the continuum limit.

(2) One has to fine tune the bare quark mass since there is an
additive quark mass renormalization.

(3) Operators of different chiral representations get mixed, etc.

\subsection{Locality}
One of the conditions of the no-go theorem was locality.
The natural definition (used here) is that the Dirac operator
in an arbitrary gauge field background falls off exponentially
i.e. satisfies the bound
\begin{equation}
  \label{bD}
  || D(x,y;U)|| \le C {\rm e}^{-\gamma |x-y|} \,,
\end{equation}
where $C$ and $\gamma>0$ are independent of the gauge field $U$.
In addition, it is also assumed that $D(x,y;U)$ depends negligibly
on $U$'s far apart from $x$ and $y$, i.e. that
\begin{equation}
  \label{bdD}
  \frac{\delta}{\delta U_\mu(z)} D(x,y;U)
\end{equation}
also falls exponentially in $|x-z|$, $|y-z|$ and $|x-y|$.
It is generally assumed that such interactions satisfy universality.

Note that it is physically too restrictive to consider only nearest
neighbour interactions or even those with strictly finite support.
In a real system the interaction coefficients die away exponentially
rather then having a finite number of strictly nonzero terms.

On the other hand, interactions decreasing more slowly then exponentially
should be considered as non-local and non-acceptable,
since universality is not expected to hold for these.
Some non-local interaction can give the same correlations
as a local one (e.g. when a bad RG transformation is used, or when
some physical fields are integrated out). This is, however, no
sufficient excuse to use them -- a slight change in the interaction
could modify the picture drastically.

Theoretically any decay rate $\gamma$ in eq.~(\ref{bD}) is allowed.
In practice, however, one works at finite lattice spacing and
$\gamma > ma$ must hold. For that reason it is important to have
a sufficiently small interaction range.

In principle it could happen that the decay rate of $D(x,y;U)$ is 
not uniformly bounded by an exponential, but there exists a bound
$\gamma(U)$ (with $\inf_U \gamma(U)=0$). In general, such interactions
should be considered non-local. However, it may happen that those
configurations when $\gamma(U) < \gamma_0$ are so strongly suppressed
that they practically `never' occur in a MC simulation.
This is a potentially dangerous possibility, but could perhaps be 
accepted with extra caution.

\section{The Ginsparg-Wilson relation}

Back in 1982 Ginsparg and Wilson \cite{GW} suggested a way to avoid
the no-go theorem and preserve consequences of the chiral symmetry.
They suggested to require instead of relation 
$ \gamma_5 D + D \gamma_5 =0$  the following milder condition
\begin{equation}
\label{1}
\gamma_5 D^{-1} + D^{-1} \gamma_5 = a 2R \gamma_5   .
\end{equation}
Here $a$ is the lattice spacing and $R$ is a {\em local} operator.
For simplicity it will be assumed also that $R$ is trivial in Dirac
indices. That $R$ commutes with $\gamma_5$ follows from eq.~(\ref{1}).
The locality of $R$ expresses the requirement that the propagating
states are effectively chiral, so at distances larger than the range
of $R$ its presence is not felt. Obviously, this is a highly
nontrivial condition since the propagator $D^{-1}$ on the lhs. is 
non-local.
(The coefficient 2 on the rhs. is for historical reasons only.)

Accordingly, the Dirac operator $D$ should satisfy 
the Ginsparg-Wilson relation
\begin{equation}
\label{2}
\gamma_5 D + D \gamma_5 = a D 2R \gamma_5 D
\end{equation}

In Ref.~\cite{GW} this relation originates from RG considerations
applied to the free fermionic action and has been generalized
to fermions in a gauge field. The local operator $R$ comes
from a chiral symmetry breaking term in the {\em block transformation}
only, not from the original action of the RG procedure.
For that reason the physics described by $D$
is expected to be chirally invariant.
(For a chirally symmetric block transformation $D$ becomes chirally 
invariant but non-local, in accordance with the no-go theorem.)

Two comments are in order here:

(1) The naive commutation property for $D$ is recovered in
the continuum limit $a\to 0$.

(2) The rhs. of eq.~(\ref{2}) is zero on solutions, i.e. for
 $D\psi=0$.

Ginsparg and Wilson suggested that a solution to eq.~(\ref{2})
is the {\em mildest} way to break chiral invariance: it should
maintain the physical consequences of chiral invariance
(soft pion theorem, etc.). However, no solution was found
for the interacting case (QCD), and the paper has been 
practically forgotten for 15 years!

Last year Peter Hasenfratz realized \cite{Hproc} that the fixed 
point (FP) action (or classically perfect action) for QCD 
\cite{QCDFP} satisfies the GW relation. 
This observation revived the interest in this relation.

\subsection{The fermionic FP action}

Let me recall the definition of the FP action.
For the pure gauge part of the action one has the saddle point
(i.e. classical) equation
\begin{equation}
\label{3}
S_{\rm g}^{\rm FP}(V)= 
\min_U \left\{S_{\rm g}^{\rm FP}(U) + T_{\rm g}(V,U) \right\} \,.
\end{equation}
Here the gauge field $V$ lives on the coarse lattice, $U$ on
the fine one, and $T_{\rm g}(V,U)$ is the blocking kernel.
This is a recursive definition, but its recursive evaluation
converges rapidly since the minimizing field $U=\overline{U}(V)$
is much smoother than the original $V$.

In the classical approximation the fermionic part remains quadratic,
and the corresponding propagator satisfies the recursion relation
\begin{eqnarray}
\label{4}
& & D^{-1}_{\rm FP}(n,n';V) =  \frac{1}{\kappa}\delta_{nn'} \\
& &~~~+\sum_{x,y} \omega(x,n;U) D^{-1}_{\rm FP}(x,y;U)\omega(y,n';U) 
\nonumber
\end{eqnarray}
Here $n,n'$ label the sites on the coarse lattice, $x,y$ on the fine
lattice, while on the rhs. $U=\overline{U}(V)$, the minimizing
configuration of the pure gauge problem is taken.
The term proportional to $1/\kappa$ comes from the chiral
breaking term in the blocking kernel for fermions; $\omega(x,n;U)$
characterizes the averaging procedure for the fermions in RG
transformation -- it is assumed to be trivial in Dirac indices
and usually it only  extends over the hypercube.
Applying eq.~(\ref{4}) recursively and using the fact that the
original $D$ (in the continuum) anti-commutes with $\gamma_5$,
it is easy to see that $D^{-1}_{\rm FP}$ satisfies eq.~(\ref{1})
with
\begin{equation}
R(U)=\frac{1}{\kappa} \left( {\rm local~operator} \right) \,.
\end{equation}
(In fact, for the RG transformation used, 
$R_{nn'} \propto \delta_{nn'}$ or it lives on the hypercube,
i.e. is even ultra-local.)

In Ref.~\cite{H2} Hasenfratz has demonstrated that, indeed,
the GW relation implies the chiral Ward identities.
The argument is the following: When the symmetry breaking part
appears in some expressions (integrated out in fermionic fields)
\begin{equation}
\label{6}
\left\langle \ldots \overline{\psi} 
DR\gamma_5 D \psi \ldots \right\rangle
\end{equation}
then the $D^{-1}$ factors from contractions of the fermionic fields
are cancelled by the $D$'s on both sides in the breaking term,
leaving only the local operator $R$. As a consequence, the symmetry
relation $\partial_\mu J_{5\mu}=0$ is satisfied up to a contact term
-- i.e. one obtains the Ward identities.
In this paper it has also been shown that the proper order parameter
for spontaneous chiral symmetry breaking is given by
\begin{equation}
\label{7}
\left\langle \overline{\psi}_x\psi_x\right\rangle_{\rm sub}\equiv
\left\langle \overline{\psi}_x\psi_x - R_{xx}\right\rangle \,.
\end{equation}
It has been proven as well that the current is not renormalized
($Z_A=Z_V=1$) and operators in different chiral representations
do not mix.
I shall discuss these issues later, in the light of newer developments.

In Ref.~\cite{LHN} it has been shown that there is an exact
Atiyah-Singer index theorem on the lattice, and the fermionic spectrum
(in arbitrary gauge background) has nice chiral properties.
The discussion was based on the FP action, but it used mostly only the
GW relation to establish these facts.

As mentioned in the introduction, the domain wall and the related 
overlap formalism also produced Dirac operators with nice chiral
properties. Neuberger \cite{NeubGW} has pointed out that the
Dirac operator obtained earlier \cite{Neub97} also satisfies
the GW relation.
To be on more general ground, I discuss first the overlap Dirac
operator. 

\subsection{Neuberger's Dirac operator}

This is an explicit construction. Introduce \cite{Neub97}
\begin{equation}
\label{8}
A=1-aD_{\rm w} \,,
\end{equation}
where $D_{\rm w}$ is the standard massless Wilson-Dirac operator 
given by eq.~(\ref{Dw}).
Note that
\begin{equation} 
\label{10}
\gamma_5 D_{\rm w} \gamma_5 =D_{\rm w}^\dagger \,.
\end{equation}
Then define
\begin{equation}
\label{11}
D=\frac{1}{a}\left( 1 -A\frac{1}{ \sqrt{ A^\dagger A}} \right)\,.
\end{equation}
It satisfies the GW relation
\begin{equation}
\label{12}
\gamma_5 D + D \gamma_5 = a D \gamma_5 D
\end{equation}
(i.e. eq.~(\ref{2}) with $R=1/2$; sometimes this special case will be
referred to as the GW relation).
Introducing $V$ through
\begin{equation}
\label{14a}
V=1-aD \,,
\end{equation}
together with the relation $\gamma_5 D \gamma_5 =D^\dagger$
one concludes that eq.~(\ref{12}) is equivalent to
\begin{equation}
\label{13}
V^\dagger V = 1 \,,
\end{equation}
i.e. $V$ is unitary.
In other words, 
\begin{equation}
\label{14}
D=\frac{1}{a}(1-V)
\end{equation}
with a unitary $V$ defines a general solution to eq.~(\ref{12}).
Of course, eq.~(\ref{14}) in itself is not enough to define an
acceptable Dirac operator. For the free case the overlap Dirac
operator has the properties (a)-(c) listed in the no-go theorem.

It will be argued later that $D$ remains local in a gauge field 
background as well.
As a consequence, $D(x,y,U)$ defines an {\em acceptable} lattice
Dirac operator (properties (a)-(c) of no-go theorem). 

Let me make a few simple comments. (Some of these straightforward
observations in this or modified form have been made independently
by several people and I shall not always give a reference.
See, however, Refs.~\cite{Bietenholz,Chiu}.)

(1) Instead of $D_{\rm w}$ in eq.~(\ref{8}) one could start
with any acceptable Dirac operator $D_0$ (massless, no doublers,
local).

(2) If the original $D_0$ satisfies the GW relation eq.~(\ref{12})
then $A^\dagger A=1$ hence $D=D_0$, i.e. the operator reproduces
itself.

(3) $H=\gamma_5 A=\gamma_5(1-D_{\rm w})$ is hermitian 
and $A^\dagger A=H^2$. As a consequence
\begin{equation}
\label{14b}
D=\frac{1}{a}\left( 1 - \gamma_5 \epsilon(H)\right)\,,
\end{equation}
where $\epsilon(x)$ is the sign function.

(4) For $A=A(\mu)=\mu-aD_{\rm w}$ (where $0<\mu<2$)
one has
\begin{equation}
\label{15}
D=\mu\frac{1}{a}\left( 1 -A\frac{1}{ \sqrt{ A^\dagger A}} \right)\,.
\end{equation}
which satisfies a slightly modified relation
\begin{equation}
\label{15a}
\gamma_5 D + D \gamma_5 = \frac{1}{\mu} a D \gamma_5 D
\end{equation}
The choice of $\mu\ne 1$ will be useful as discussed later.
Note that for $\mu <0$ there are no massless fermions at all,
while for $\mu >2$ there are too many, since the modes around
$\lambda(D_{\rm w})=2$ (the massive doublers) also become
massless modes of $D$. Therefore only $0<\mu<2$ produces an acceptable
Dirac operator.

(5) It is also easy to generalize to the case of arbitrary local
$R$:
\begin{equation}
\label{16}
A=1-\sqrt{2R} \, D_0 \sqrt{2R} \,,
\end{equation}
and
\begin{equation}
\label{16a}
D=\frac{1}{a}\frac{1}{\sqrt{2R}}
\left( 1 -A\frac{1}{ \sqrt{ A^\dagger A}} \right)
\frac{1}{\sqrt{2R}}\,.
\end{equation}
This satisfies eq.~(\ref{2}). For simplicity, however, we shall
consider mostly the case $R=1/2$.

\section{The index theorem on the lattice}
For a massless Dirac operator in the continuum theory the Atiyah-Singer 
index theorem \cite{AS} holds:
\begin{equation}
\label{17}
Q_{\rm top}={\rm index}(D) \,.
\end{equation}
Here 
\begin{equation}
\label{18}
Q_{\rm top}=\frac{1}{32\pi^2}\int d^4x \epsilon_{\mu\nu\rho\sigma}
{\rm tr}\left( F_{\mu\nu}F_{\rho\sigma}\right)
\end{equation}
is the topological charge of the gauge field, and
\begin{equation}
{\rm index}(D)\equiv n_- - n_+ \,,
\end{equation}
where $n_+$, $n_-$ is the number of solutions to $Du=0$ with $+/-$
chiralities, i.e. $\gamma_5 u = \pm u$.

The spectrum of $D$ in the continuum consists of
\begin{itemize}
\item $\lambda=0$; the corresponding modes have definite chirality
\item $\lambda=\pm i \alpha$ ($\alpha$ real, non-zero); for the
corresponding modes $(u_\lambda,\gamma_5 u_\lambda)=0$.
\end{itemize}
Obviously, $n_- - n_+$ is topological invariant: an unpaired
eigenvalue
at $\lambda=0$ cannot be continuously produced or destroyed.

In the lattice formulation one has problems with both sides 
of eq.~(\ref{17}).
On the lhs. there are several ways to assign $Q_{\rm top}(U)$
to a given gauge configuration (field theoretical definition,
geometrical definition). To avoid ambiguities for strongly
fluctuating gauge fields,  cooling is often applied.
An additional problem is that, in general, dislocations could be
present, i.e. configurations which violate the inequality
\begin{equation}
\label{20}
S_{\rm g}(U) \ge S_{\rm inst}^{\rm cont} \cdot |Q_{\rm top}(U)| \,.
\end{equation}

On the rhs. in an arbitrary gauge field $D_{\rm w}$ has no exactly
zero eigenvalues -- they are distorted by lattice artifacts.
The spectrum of $D_{\rm w}$ consists of
\begin{itemize}
\item real eigenvalues, with $(u_\lambda,\gamma_5 u_\lambda)\ne 0$
but not exactly $\pm 1$
\item complex eigenvalues, in cc. pairs, $\lambda$ and $\lambda^*$
with $(u_\lambda,\gamma_5 u_\lambda)=0$.
\end{itemize}

For smooth instanton background ($\rho \gg a$) $D_{\rm w}$ has 
an isolated real, nearly zero eigenvalue with
$(u_\lambda,\gamma_5 u_\lambda) \approx +1$ or $-1$.
However, for a small instanton, or a typical MC configuration
(say at $\beta=6.0$) there are a lot of small real eigenvalues.
Therefore the definition of topological charge based on investigating
the spectrum of $D_{\rm w}$ is quite ambiguous \cite{SV,Itoh}.

Let me remind you that the RG approach has a theoretically appealing
(although  difficult in practice) way to define the topological charge
of a given configuration, in spin models \cite{BBHN} and in gauge
theories \cite{DHZK}. The solution is the following:
take the minimizing configuration on the fine lattice
\begin{equation}
\label{21}
U_{\rm fine}=\overline{U}(U_{\rm coarse})
\end{equation}
and define the FP topological charge on $U_{\rm coarse}$ through
\begin{equation}
\label{22}
Q_{\rm top}^{\rm FP}(U_{\rm coarse}) = 
Q_{\rm top}^{\rm naive}(U_{\rm fine}) \,. 
\end{equation}
It is easy to show that with this definition there are no
dislocations, i.e. 
\begin{equation}
\label{23}
S^{\rm FP}(U) \ge 
S_{\rm inst}^{\rm cont} \cdot | Q_{\rm top}^{\rm FP}(U)| 
\end{equation}
holds for any configuration.

As has been shown in Ref.~\cite{LHN} the FP fermion operator has nice
chiral properties which allow one to establish the exact index theorem
on the lattice. 

Let me discuss the main points. The GW relation eq.~(\ref{12})
(with $a=1$) and $\gamma_5 D \gamma_5 = D^\dagger$ implies
\begin{equation}
\label{25}
D+D^\dagger=D D^\dagger=D^\dagger D \,,
\end{equation}
hence $D$ and $D^\dagger$ commute, i.e. $D$ is normal.
(As a consequence, the eigenvectors corresponding to different
eigenvalues are orthogonal to each other.) 
The spectrum of $D$ (cf. eq.~(\ref{14})) is on the circle 
$\lambda=1-{\rm e}^{i\alpha}$.
There are three types of eigenvalues
\begin{itemize}
\item $\lambda=0$, with $\gamma_5 u_\lambda=\pm u_\lambda$
(denote their numbers by $n_+$, $n_-$),
\item $\lambda=2$, with $\gamma_5 u_\lambda=\pm u_\lambda$
(with multiplicities  $n_+'$, $n_-'$),
\item $\lambda=$ complex, with $\gamma_5 u_\lambda=u_{\lambda^*}$
and $(u_\lambda,\gamma_5 u_\lambda)=0.$
\end{itemize}
Note that \cite{Chiu} since ${\rm Tr}(\gamma_5)=0$ the index of
$D$ and $2-D$ is opposite:
\begin{equation}
\label{26}
n_+' - n_-' = n_- - n_+ \,.
\end{equation}

Define the topological charge density by
\begin{equation}
\label{27}
q(x)=\frac{1}{2} {\rm tr} \left( \gamma_5 D(x,x)\right).
\end{equation}
For the topological charge $Q_{\rm top}= \sum_x q(x)$ we have
\begin{eqnarray}
Q_{\rm top} &=& \frac{1}{2} {\rm Tr}(\gamma_5 D) 
= -\frac{1}{2} {\rm Tr}(\gamma_5 (2-D)) \nonumber \\
&=& -\frac{1}{2} \sum_\lambda 
(2-\lambda)(u_\lambda,\gamma_5 u_\lambda) \label{28}\\
&=& n_- - n_+ \equiv {\rm index}(D) \nonumber \,.
\end{eqnarray}
Here we have used the properties of eigenvectors of $D$ listed above.

Obviously, $Q_{\rm top}$ is topological invariant since the integer
$n_- - n_+$ cannot change smoothly. This fact can also be obtained
directly from the GW relation.

Equation (\ref{28}) is the index theorem on the lattice: it states that
one can define a (gauge invariant, real) topological charge density
$q(x;U)$ on a given gauge configuration which leads to an integer
topological charge. The topological charge defined this way equals to
the index of $D$. Both sides are defined through $D$.
Is this a tautology? No: the important point here is that
one has a local topological charge density not just an integer
number defining the charge.

It is easy to see that for smooth configurations
\begin{equation}
\label{29}
q(x)=
\frac{1}{32\pi^2}\epsilon_{\mu\nu\rho\sigma}
{\rm tr}\left( F_{\mu\nu}F_{\rho\sigma}\right) + O(a^2)
\end{equation}
since $q(x)$ is a gauge invariant, pseudoscalar, dimension 4 operator
and the prefactor is fixed by the fact that in the continuum limit
$a\to 0$ one should recover the continuum topological charge.

I would like to stress that {\em any acceptable} solution to the GW
relation defines a topological charge for which the exact index
theorem holds. Note, however, that different solutions will give,
in general, different values for the topological charge
 -- they need to agree only for sufficiently smooth configurations. 
 From a practical point of view the operator
$D$ has to provide a topological charge which is robust enough,
giving a reasonable answer also for relatively rough configurations.
To illustrate this point, consider Neuberger's operator in
eq.~(\ref{15}) with a small $\mu$, say $\mu=0.01$. For this choice
only very large instantons are counted as such when defined through
eq.~(\ref{28}) -- perhaps only those with $\rho/a > 100$.
Smaller instantons ``fall through the lattice'' since the smallest
eigenvalue of $D_{\rm w}$ would shift beyond $\mu$.
Obviously, this would be a bad Dirac operator and a bad topological
charge -- in spite of the fact that they satisfy the exact index
theorem, eq.~(\ref{28}).

Any {\em acceptable} discretization of the Dirac operator
(without requiring to be a solution to the GW relation)
could be used to determine the correct topological charge
for {\em sufficiently small} lattice spacing $a$ -- the exactly 
zero eigenvalue of the continuum Dirac operator will move
only slightly away from zero, remaining real, and no other 
nearly zero eigenvalues will appear. The extra requirement of
GW relation forbids the eigenvalue to move away from zero
for such configurations.
Hence no extra conditions (beyond acceptability and GW relation)
are required for the index theorem to hold. 
However, as pointed out above, these conditions do not guarantee
that the topological charge defined this way is the `correct' one
-- except for sufficiently large instantons. Of course, there is
no unique way to define the topological charge on the lattice,
but for not-too-small instantons (with a radius of a few lattice
spacing, say) the geometrical definition is natural. For practical
reasons, one would like to have a GW Dirac operator $D$
which reproduces the natural topological charge of not too large
instantons as well.
I discuss these points perhaps more than necessary, because there
is some controversy in the literature \cite{Chiu,ChiuZenkin}.

In Ref.~\cite{LHN} it has also been shown that for the FP Dirac
operator one has in addition
\begin{equation}
\label{30}
Q_{\rm top}^{\rm gaugeFP}=
\sum_x {\rm tr}\left( \gamma_5 R D^{\rm FP}\right)_{xx}
\equiv Q_{\rm top}^{\rm fermFP}\,,
\end{equation}
i.e. the topological charge defined by $D^{\rm FP}$ {\em coincides}
with the old definition in pure gauge theory. In this case, for
example, there are no dislocations, and one expects small lattice
artifacts. 

An important point is that $D(x,y;U)$ is discontinuous in the gauge
field. The reason is that on the lattice one can connect continuously
any two configurations, but ${\rm index}(D(U))$ has a jump on the path
connecting, say, $Q_{\rm top}=0$ and 1.

With the FP action the origin of this discontinuity is that
the minimizing configuration $\overline{U}(V)$ becomes ambiguous
when the rhs. of eq.~(\ref{3}) has a degenerate minimum
(when the instanton ``falls through the lattice'').
For Neuberger's Dirac operator the discontinuity appears
when a real eigenvalue of $D_{\rm w}$ passes 1.
In this case an eigenvalue of $H=\gamma_5 (1-D_{\rm w})$ 
changes sign, and $\epsilon(H)$ in eq.~(\ref{14b}) is discontinuous.

Note that a similar discontinuity appears when the fermions are 
defined in the continuum, in a continuous gauge background obtained
by interpolating the lattice gauge field \cite{tHooft}.
In this case the process of interpolation becomes discontinuous,
similarly to the situation with the FP action.

An important question: does $D$ become non-local when this
discontinuity occurs? This would be a very bad news for the 
present approach. Fortunately, this is not true in general.
In the FP action the two solutions $U_1$ and $U_2$ corresponding 
to the degenerate minima are expected to differ only locally.
Neuberger's $D$, as shown by L\"uscher \cite{loc,HJL}, remains 
also local when an isolated eigenvalue of $H$ passes zero, 
so a discontinuity of D does not imply a loss of locality.

\section{The chiral condensate}

Restoring the lattice spacing $a$, the spectrum of $D$ is given 
by a circle $\lambda=\frac{1}{a}\left(1-{\rm e}^{i\alpha}\right)$
with $-\pi \le \alpha < \pi$, crossing the real axis at $\lambda=0$
and $\lambda=2/a$. Correspondingly, the spectrum of $D^{-1}$
is a straight line parallel to the imaginary axis, of the form
$1/\lambda=a/2+i\gamma$, $-\infty < \gamma < +\infty$ for arbitrary
gauge field.

Denote the unnormalized expectation value of 
${\cal O}(\overline{\psi},\psi)$ in a fixed gauge background by
\begin{equation}
\label{31}
\left\langle {\cal O}(\overline{\psi},\psi) \right\rangle_{\rm F}
\equiv \int d\overline{\psi} d\psi 
{\rm e}^{-S_{\rm F}(\overline{\psi},\psi) } 
{\cal O}(\overline{\psi},\psi) \,.
\end{equation}
Then we have
\begin{equation}
\label{32}
a^4\sum_x \left\langle \overline{\psi}_x \psi_x \right\rangle_{\rm F}=
\left\langle 1 \right\rangle_{\rm F} \sum_\lambda \frac{1}{\lambda} \,.
\end{equation}
Obviously, the $a/2$ part of $1/\lambda$ is independent of
the dynamics and has to be subtracted to define a proper chiral 
order parameter. This yields to
\begin{equation}
\label{33}
\left\langle \overline{\psi}_x \psi_x \right\rangle_{\rm sub}=
\left\langle \overline{\psi}_x \psi_x -
2N_{\rm f}N_{\rm c}\frac{1}{a^3} \right\rangle \,,
\end{equation}
the same as suggested in Ref.~\cite{H2}.

Adding a mass term $m$ to $D$ would shift the whole circle of
eigenvalues.
Since the real eigenvalues at $\lambda=2/a$ also contribute
to chiral relations it is better to leave them unchanged.
This is simply achieved by the replacement
\begin{equation}
\label{34}
D \to \left( 1 - \frac{1}{2}am\right) D + m
\end{equation}
which not only shifts the circle but also rescales its radius
appropriately.
With this definition of $m$ one obtains
\begin{equation}
\label{34a}
\frac{\partial}{\partial m}\langle 1 \rangle_{\rm F}
= \left\langle \overline{\psi}\left( 1 - \frac{1}{2}aD\right)
\psi \right\rangle_{\rm F}\,.
\end{equation}
This expression is identical to eq.~(\ref{33}), i.e. the proper
order parameter. This equivalent form on the rhs. has been suggested
first by Chandrasekharan \cite{Chand}, who obtained it from 
L\"uscher's exact symmetry.

\section{Exact chiral symmetry on the lattice}

At this moment we are in a peculiar situation: all signs
of a symmetry are present (Ward identities, index theorem, 
chiral spectrum, \ldots) but the action is not chirally invariant.

L\"uscher made an important observation that in fact the action
possesses an exact symmetry.
Perform an infinitesimal change of variables
$\psi \to \psi + i\epsilon\delta\psi$ and 
$\overline{\psi} \to \overline{\psi} 
+ i\epsilon\delta\overline{\psi}$ with a global flavour singlet
transformation 
\begin{eqnarray}
\delta\psi &=& \gamma_5 \left( 1-\frac{1}{2}aD\right)\psi \label{35} \\
\delta\overline{\psi} &=& 
\overline{\psi}\left( 1-\frac{1}{2}aD\right)\gamma_5  \nonumber
\end{eqnarray}
 From the GW relation eq.~(\ref{12}) it follows that the action is
invariant
\begin{equation}
\delta(\overline{\psi} D \psi )=0 \,.
\end{equation}
The corresponding flavour non-singlet transformation is given by
\begin{eqnarray}
\delta\psi &=& T \gamma_5 \left( 1-\frac{1}{2}aD\right)\psi \label{36} \\
\delta\overline{\psi} &=& 
\overline{\psi}\left( 1-\frac{1}{2}aD\right)\gamma_5 T  \nonumber
\end{eqnarray}

Now it seems that we have more symmetry than possible --
the flavour singlet transformation should be anomalous. 
Where is the anomaly hidden? As pointed out by L\"uscher,
the fermionic integration measure is {\em not} invariant
under the singlet transformation in eq.~(\ref{35}) 
\begin{eqnarray}
\delta[d\overline{\psi}d\psi] &=& {\rm Tr}(a\gamma_5 D)
[d\overline{\psi}d\psi] \label{37} \\
 &=& 2N_{\rm f}(n_- - n_+)[d\overline{\psi}d\psi] \nonumber\,.
\end{eqnarray}
That is, the measure breaks the flavour singlet chiral symmetry
in a topologically non-trivial gauge field where 
${\rm index}(D)= n_- - n_+ \ne 0$.
The corresponding (global) Ward identity  for the flavour singlet
transformation is
\begin{equation}
\label{38}
\left\langle \delta{\cal O}\right\rangle_{\rm F} = 
2N_{\rm f} \nu \left\langle {\cal O}\right\rangle_{\rm F} \,,
\end{equation}
where $\nu = Q_{\rm top}(U)$.

Note that the non-singlet symmetry is not anomalous,
$\left\langle \delta{\cal O}\right\rangle_{\rm F} = 0$ since eq.~(\ref{37})
for this case contains ${\rm Tr}(a\gamma_5 DT)\propto {\rm tr}(T)=0$.

Using these exact symmetries of the action, the Ward identities
follow, of course, directly. By taking
\begin{equation}
\label{40}
{\cal O} = \sum_x \overline{\psi}_x t_a\gamma_5 \psi_x \,,
\end{equation}
and a non-singlet transformation one has 
$\left\langle \delta{\cal O}\right\rangle_{\rm F} = 0$ i.e. \cite{Chand}
\begin{equation}
\label{41}
\left\langle \overline{\psi}\left( 1 - \frac{1}{2}aD\right)
\psi \right\rangle_{\rm F} = 0 ~~~(m=0, ~ V ~\mbox{finite})\,.
\end{equation}
This is the order parameter which is expected to be broken
spontaneously in the thermodynamic limit \cite{H2}:
\begin{equation}
\label{42}
\lim_{m\to 0} \lim_{V\to\infty}
\left\langle \overline{\psi}\left( 1 - \frac{1}{2}aD\right)
\psi \right\rangle = \Sigma \ne 0 \,.
\end{equation}
In Ref.~\cite{Chand} there is also a relation connected with the
U(1) problem (massive $\eta'$ meson) -- this will be discussed later.

Whether the chiral symmetry is indeed broken spontaneously in QCD
is a delicate dynamical question. The chiral invariant
Dirac operator provides a firm basis to investigate this question
by avoiding an explicit breaking.

Note that the Nielsen-Ninomiya no-go theorem is avoided here
by the non-standard form of the lattice chiral symmetry:
$D$ is not invariant under the naive $\gamma_5$ symmetry,
nevertheless it is invariant under the generalized transformation.

\section{Chiral decomposition}

In the continuum, chiral symmetry allows one to decompose the fermion 
field into left and right chiral components, transforming
independently. Besides yielding a convenient formalism for vector
theories like QCD, this decomposition is a necessary first step 
towards the chiral gauge theories.

The matrix $\gamma_5$ anti-commutes with the continuum Dirac operator
and $\gamma_5^2=1$. This allows to introduce the chiral projectors
\begin{equation}
\label{43}
P_\pm = \frac{1}{2}(1\pm \gamma_5) \,,
\end{equation}
and the action decomposes as 
$D_{\rm cont} = P_+ D_{\rm cont} P_- + P_- D_{\rm cont} P_+$ .

On the lattice it is useful to introduce \cite{HNu,L-ChGT}
the operator 
\begin{equation}
\label{44}
\hat{\gamma_5}=\gamma_5 (1-aD)=\gamma_5 V \,,
\end{equation}
where $D$ satisfies the GW relation eq.~(\ref{12}).
Remember that $V$ is unitary and $\gamma_5 V\gamma_5 =V^\dagger$.
Then 
\begin{equation}
\label{45}
\gamma_5 D = -D \hat{\gamma}_5 \,, \mbox{~~and~~}\hat{\gamma}_5^2 = 1 \,.
\end{equation}
One can introduce the projectors
\begin{equation}
\label{47}
\hat{P}_\pm = \frac{1}{2}\left( 1 \pm \hat{\gamma}_5 \right) \,.
\end{equation}
They satisfy the relations
\begin{equation}
\label{48}
D \hat{P}_+ = P_- D \,, ~~ D \hat{P}_- = P_+ D \,,
\end{equation}
and allow the decomposition
\begin{equation}
\label{49}
D= P_+ D \hat{P}_- + P_- D \hat{P}_+ \,.
\end{equation}
To conclude the definitions let us introduce the chiral
components of $\psi$:
\begin{eqnarray}
& & \psi_L= \hat{P}_- \psi \,, ~~~  
\overline{\psi}_L= \overline{\psi} {P}_+ \,, \label{50}\\
& & \psi_R= \hat{P}_+ \psi \,, ~~~  
\overline{\psi}_R= \overline{\psi} {P}_- \,. \label{51}
\end{eqnarray}
In terms of these components we have
\begin{equation}
\label{52}
\overline{\psi} D \psi= \overline{\psi}_L D \psi_L +
 \overline{\psi}_R D \psi_R \,,
\end{equation}
analogously to the continuum case.
Note that eqs.~(\ref{50},\ref{51}) are not symmetric in 
$\psi$ and $\overline{\psi}$. As discussed later, this
asymmetric definition seems to play an essential role.

Analogously to the continuum case, the mass term should
couple the L and R components (but not the L,L or R,R
components):
\begin{equation}
\label{53}
{\cal S} \equiv \overline{\psi}_L \psi_R + \overline{\psi}_R \psi_L
= \overline{\psi} \left( 1-\frac{1}{2}aD\right) \psi \,.
\end{equation}
The corresponding pseudoscalar is given by
\begin{equation}
\label{54}
{\cal P} \equiv \overline{\psi}_L \psi_R - \overline{\psi}_R \psi_L
= \overline{\psi} \gamma_5 \left( 1-\frac{1}{2}aD\right) \psi \,.
\end{equation}
The scalar (pseudoscalar) densities ${\cal S}$ and ${\cal P}$
will often appear in the formulae below.
 
Performing a left chiral transformation means
\begin{eqnarray}
& & \delta \psi_L = -\psi_L\,,~~~ 
 \delta \overline{\psi}_L = \overline{\psi}_L \,, \label{55} \\
& & \delta \psi_R = 0\,,~~~ 
 \delta \overline{\psi}_R = 0 \,, \label{55a}
\end{eqnarray}
Using this transformation one has
\begin{equation}
\label{56}
\delta{\cal P}={\cal S} \,,~~~ \delta{\cal S}={\cal P} \,.
\end{equation}

The combinations in eqs.~(\ref{53},\ref{54}) are the natural
generalizations of the corresponding continuum expressions.
Note that the $1-\frac{1}{2}aD$ factor cancels the contribution
of the unphysical real eigenmode at $\lambda=2/a$.
Using these combinations of fields, one obtains very simple,
compact relations, for example:
\begin{eqnarray}
& & \sum_x \left\langle {\cal P}_x \right\rangle_{\rm F} =  \label{57} \\
& & - N_{\rm f} {\rm Tr} \left( \gamma_5 
\frac{\left(1-\frac{1}{2}aD\right)}{\left(1-\frac{1}{2}am\right)D+m}
\right)
\cdot \left\langle 1 \right\rangle_{\rm F} = \nonumber \\
& & - \frac{1}{m}N_{\rm f} (n_+-n_-) \left\langle 1 \right\rangle_{\rm F} =
\frac{1}{m}N_{\rm f} \nu \left\langle 1 \right\rangle_{\rm F} \nonumber \,.
\end{eqnarray}
Note that only the $\lambda=0$ eigenmodes contribute when evaluating
the trace.

It is instructive to consider the flavour singlet Ward identity
which leads to a relation for the $\eta'$ propagator:
\begin{equation}
\label{58}
\frac{m}{V}\sum_{x,y} 
\left\langle {\cal P}_x {\cal P}_y \right\rangle_{\rm F}
= \frac{1}{V}\sum_x \left\langle {\cal S}_x \right\rangle_{\rm _F}
+ \frac{N_{\rm f}^2}{m}\frac{\nu^2}{V} \langle 1 \rangle_{\rm F} \,.
\end{equation}
If the $\eta'$ stays massive in the chiral limit $m\to 0$ then
the lhs. (after averaging over the gauge fields) should go to zero.
This leads to the interesting relation \cite{Chand} between the
topological susceptibility and chiral condensate in full QCD
\begin{equation}
\label{59}
\frac{1}{V}\left\langle \nu^2\right\rangle =
-\frac{m}{N_{\rm f}^2}\left\langle {\cal S} \right\rangle
+{\rm O}(m^2) \,.
\end{equation}
This relation has been obtained previously in the continuum 
\cite{LS,VV} but it is important to have its lattice version
as well. It shows, for example that the topological susceptibility
must be a well behaved physical quantity on the lattice.

For the flavour non-singlet variant of eq.~(\ref{58}) the term
proportional to $\nu^2$ is absent since this symmetry is not
anomalous. Therefore a non-vanishing chiral condensate,
eq.~(\ref{42}), yields to (quasi) Goldstone bosons with mass
$m_\pi^2 \propto m\Sigma$.

We conclude this section with a few remarks. 
Note that $\hat{\gamma}_5$ coincides with $\epsilon(H)$ introduced
in the overlap formalism.
Another point worth mentioning is that for the modes with
$\lambda=0$ one has $\hat{\gamma}_5=\gamma_5$ while for the
modes at $\lambda=2/a$ they differ in sign, 
$\hat{\gamma}_5=-\gamma_5$, as seen from the definition
eq.~(\ref{44}). This observation will be important when
we consider chiral fermions.

\section{The $\theta$ parameter}

We can now introduce the $\theta$ parameter,
as in the continuum. Here I shall restrict the discussion
to some basic points only.

Define the lattice Lagrangean with a complex mass $m$ and 
the $\theta$-parameter:
\begin{equation}
\label{61}
{\cal L}= {\cal L}_{\rm g}+\overline{\psi}D\psi +
m \overline{\psi}_R \psi_L +
m^* \overline{\psi}_L \psi_R - i\theta q(x) \,,
\end{equation}
where ${\cal L}_{\rm g}$ is the gauge action and
$q(x)$ is defined through $D$ by eq.~(\ref{27}).
By changing the variables
\begin{equation}
\label{63}
\psi_L \to {\rm e}^{-i\alpha}\psi_L \,,~~~
\overline{\psi}_L \to {\rm e}^{i\alpha}\overline{\psi}_L \,.~~~
\end{equation}
one can restore the invariance of the action by 
$m\to{\rm e}^{i\alpha}m$ and the non-invariance of the measure 
can be compensated by $\theta \to \theta -N_{\rm f} \alpha$.
Therefore the free energy should depend only on the combination
$m\exp(i\theta/N_{\rm f})$, as in the continuum.
One can proceed now e.g. by repeating (now on the lattice) the steps 
of Ref.~\cite{LS} where e.g. useful sum rules have been obtained 
for the spectrum of the Dirac operator. To illustrate this, 
consider the partition function
\begin{equation}
\label{64}
Z \equiv Z(\theta) = \sum_\nu {\rm e}^{i\theta \nu} Z_\nu
= {\rm e}^{-F(m,\theta)} \,.
\end{equation}
Under some natural assumptions \cite{LS} one has for the free energy
\begin{equation}
\label{65}
F(m,\theta)= -V N_{\rm f} \Sigma {\rm Re} 
\left(m{\rm e}^{i\theta/N_{\rm f}}\right) + {\rm O}(m^2) \,.
\end{equation}
Taking then the second derivative with respect to $\theta$ 
(for real $m$) one obtains
\begin{equation}
\label{66}
\left\langle \nu^2 \right\rangle = V\Sigma m \frac{1}{N_{\rm f}}
\end{equation}
which is equivalent to eq.~(\ref{59}).

As another example, it is easy to show that the fermion determinant 
is given by
\begin{eqnarray}
& & \det D(U)= \label{67} \\
& & (2m)^\nu \prod_{\lambda}
\left( |\lambda|^2 + 
\left(1-\frac{|\lambda|^2}{4} \right) |m|^2 \right) \nonumber \,,
\end{eqnarray}
for $\nu \ge 0$, and the product is over the complex eigenvalues only.
For $\nu<0$ the prefactor is replaced by $(2m^*)^{-\nu}$.
Eq.~(\ref{67}) differs from its continuum counterpart by the
presence of the factor $1-|\lambda|^2/4$.

\section{Cut-off effects with GW Dirac operators}

An important consequence of the exact chiral symmetry on the lattice
is that there are no O($a$) lattice artifacts, to any order of $g$.
The reason is the following. The O($a$) lattice artifacts of any action
could be compensated by adding an extra term which is a lattice
version of $\overline{\psi}\sigma_{\mu\nu}F_{\mu\nu}\psi$
with an appropriately chosen coefficient \cite{LSSW}. 
The original action in our case has the exact (generalized) chiral 
symmetry of eq.~(\ref{35}) and its artifacts could be compensated 
only by a similarly invariant extra term. Any lattice regularization of 
$\overline{\psi}\sigma_{\mu\nu}F_{\mu\nu}\psi$ will violate, however,
this symmetry, hence the O($a$) artifacts should be absent for the
original lattice action as well. 
Note also that the clover term needed for O($a$) improvement 
has been determined by requiring that some Ward identity holds
independently of the gauge field. This is, however, automatically 
satisfied here for $m=0$ hence there is no need for extra clover term.

In addition, in the presence of a mass $m$ there are also no
O($am$) artifacts, if the mass is  introduced properly, as in
eq.~(\ref{61}). Then the symmetry $m\to -m$ forbids such term.
Indeed, under a finite chiral transformation
\begin{equation}
\label{67a}
\psi \to \psi'= \hat{\gamma}_5 \psi \,,~~~
\overline{\psi} \to -\overline{\psi} \gamma_5
\end{equation}
$\overline{\psi} D \psi$ is invariant while the mass term
$\overline{\psi} (1-\frac{1}{2}aD)\psi$ changes sign.
Although due to the anomaly the partition function is not invariant
for $\nu\ne 0$, this fact is not relevant for our considerations.
Indeed, for $m\ne 0$ and sufficiently large volume one can restrict 
the system to the topologically trivial sector $\nu=0$ \cite{LS}. 

Note, however, that the O($a^2$) lattice artifacts could still be
large at the relevant lattice spacings, and could be quite different
for different  solutions of the GW relation.

Finally, it is interesting to see (although it is not necessary
after the previous discussion) that the absence of O($a$) artifacts
on the tree level follows directly from the GW relation:
expanding $D$ in powers of $a$ it implies that
\begin{equation}
\label{68}
D=i\gamma_\mu\nabla_\mu+ a c \left( \nabla^2 +\frac{i}{2}
\sigma_{\mu\nu} F_{\mu\nu} \right) +{\rm O}(a^2) \,,
\end{equation}
(with some constant $c$)
i.e. a term with $c_{\rm SW}=1$ is automatically generated.

\section{Properties of Neuberger's Dirac operator}

Here we shall discuss some basic properties of Neuberger's Dirac
operator and its straightforward generalizations.

\begin{figure}[t]
\begin{center}
\vspace*{-1mm}
\epsfig{file=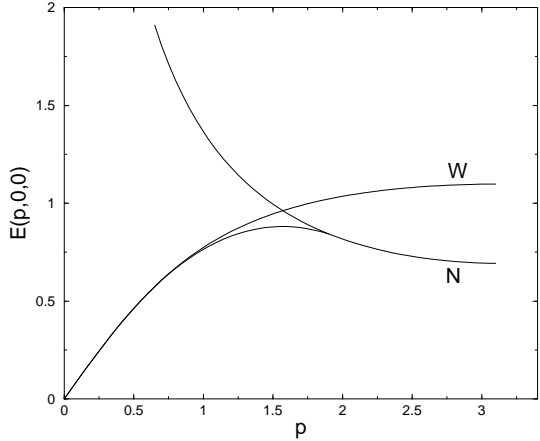,angle=-90,width=7cm}
\vspace*{-10mm}
\end{center}
\caption{\label{fig:spect}
Spectrum of the Wilson-Dirac operator (W), and of Neuberger's
operator (N).}
\end{figure}

Consider first the free case. The spectrum $E(\vec{p})$ of 
$D$ defined in eq.~(\ref{11}) is shown in fig.~\ref{fig:spect}.
together with the spectrum for the standard Wilson-Dirac operator
for $\vec{p}=(p,0,0)$.
The latter is denoted by W and given by
\begin{equation}
\label{71}
\cosh E = \frac{3-2\cos p}{2-\cos p} \,,
\end{equation}
while N denotes the two branches for Neuberger's Dirac operator
where $\cosh E(\vec{p})$ is  given by
\begin{equation}
\label{72}
\sqrt{1+\sin^2 p} \,, \mbox{~~and~~} 
\frac{3-2\cos p}{2(1-\cos p)} \,.
\end{equation}
(Note that the lower curve extends only up to the point where the two
branches join each other -- this peculiar behaviour is due to the
non-analyticity in the square root.)
The spectrum for $ap>1$ is worse than that for $D_{\rm w}$. 
This illustrates that a solution of GW relation can have large 
O($a^2$) artifacts -- the fact that $D$ is spread over some distance 
does not necessarily make the artifacts small!

The locality of $D$ for the free case is guaranteed by the
inequality
$|| D(x-y)|| \le C {\rm e}^{-\gamma |x-y|_0}$
where $|x-y|_0 = \max_\mu |x-y|_\mu$ and $\gamma=0.693\ldots$.

\begin{figure}[t]
\begin{center}
\vspace*{-1mm}
\epsfig{file=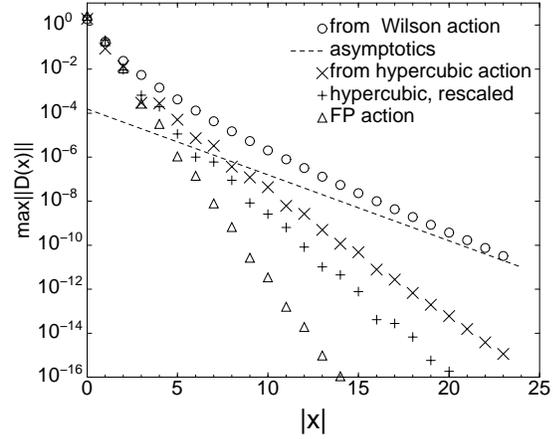,angle=-90,width=7cm} 
\vspace*{-7mm}
\end{center}
\caption{\label{fig:loc}
Locality of different GW Dirac operators for the free fermions.}
\end{figure}

This asymptotic behaviour is shown on fig.~\ref{fig:loc} 
by dashed line. 
The decay of $||D(x)||$ for Neuberger's $D$ is shown by circles. 
Two more solutions to GW relation are added. 
Crosses show the case with $A=1-D_{\rm HC}$ where $D_{\rm HC}$ 
is a FP action truncated to the hypercube. 
Since there is a non-trivial $R$ in this case the more
natural choice, $A=1-\sqrt{2R} D_{\rm HC} \sqrt{2R}$,
(cf. eqs.~(\ref{16},\ref{16a})) is also shown (+ signs). 
Finally, the true FP action is also indicated (triangles). 
It is obvious that there are rather large differences in
the range of interactions for different GW Dirac operators, and this
is an important factor in making a choice for numerical simulations
(see also Ref.~\cite{Bietenholz}).

A less trivial point is the question of locality in a gauge field
background.
In Ref.~\cite{HJL} it is proven that for small gauge fields
\begin{equation}
\label{74}
|| 1-U_{\rm pl} || < \epsilon \mbox{~~~ for all plaquettes}
\end{equation}
with $\epsilon < 1/30$, the corresponding $D(x,y;U)$ is local with
exponentially decaying tail, and the localization length is uniformly
bounded.

Although eq.~(\ref{74}) should hold sufficiently close to the
continuum limit, it is violated at practical values of $a$.
Theoretically one can restrict the gauge fields to satisfy
eq.~(\ref{74}). This is sufficient to show that there exists
an exactly chiral invariant, acceptable lattice regularization
of QCD. However, the restriction leads to an astronomical correlation
length, hence it is only of theoretical interest.
As mentioned before, in Ref.~\cite{HJL} it is also shown that
an isolated zero of $A$ does not spoil locality.

\begin{figure}[t]
\begin{center}
\vspace*{-1mm}
\epsfig{file=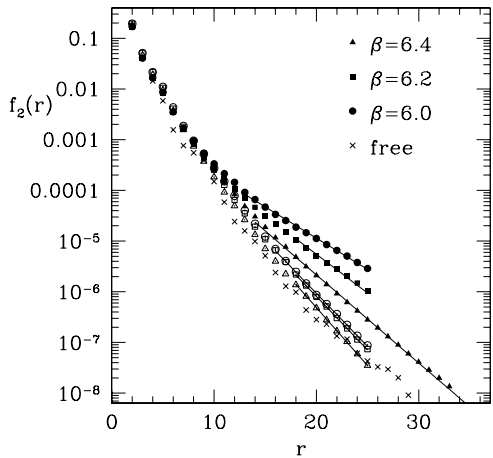,angle=0,width=7.5cm}
\vspace*{-25mm}
\end{center}
\caption{\label{fig:f2r}
Locality of Neuberger's Dirac operator in gauge field background.
}
\end{figure}

An important question is the locality in gauge fields generated
by MC simulations. First results on this are presented in
Refs.~\cite{HJL,Montvay}. In fig.~\ref{fig:f2r} results from 
Ref.~\cite{HJL} are shown, for SU(3) gauge fields generated 
at $\beta=6.4$, $6.2$ and $6.0$ (triangles, squares, circles). 
The full symbols are for $\mu=1$ while the open ones for (roughly) 
optimized values, $\mu=1.2$ for $\beta=6.4$ and  $\mu=1.4$ for 
$\beta\le 6.2$.

It is also important to have a sufficiently effective way of
evaluating the inverse square root. This issue is discussed 
in refs.~\cite{Chiu,HJL,Montvay,Bunk,Neub3,Edw,Kenn}.

I only mention here that it seems to be a feasible task
but still much has to be done in this direction.

A few comments are in order:

(1) Unlike evaluating $D^{-1}$ for massless quarks, taking the
inverse square root is not a critical problem -- appearance
of zero modes of $A$ is accidental.

(2) Starting with a good approximation of the FP action
instead of $D_{\rm w}$ (as in eq.~(\ref{16})) the probability
to have a zero mode of $A$ is strongly suppressed, since with the
exact FP action $A$ would be unitary. Note, however, that
for any continuous parametrization of the FP action, $A$ must
have a zero mode for specific configurations, when an instanton
``falls through the lattice''. This happens on the boundary
separating regions of gauge field configurations with 
different topological charges.

(3) Using a GW Dirac operator there will be no ``exceptional
configurations'' in the usual sense \cite{Exc} -- these would
correspond to eigenvalues $\lambda(D)<0$. 
(One should distinguish these from configurations where 
$\lambda(A)=0$ appearing in this approach.)

(4) One can construct, in principle, configurations when many
``instantons fall through'' simultaneously at different sites
of a large lattice. This would produce a situation when $A$ has many
nearly zero eigenvalues. Will $D$ be local on such configurations?
One expects that such configurations have negligible weight in the 
partition function. Will their effect be also negligible?
I think, these questions have to be investigated.

\section{Measuring the topological charge}

There are several equivalent ways to measure the topological
charge defined by $D$ through eqs.~(\ref{27},\ref{28}):

(1) find the zero eigenmodes of $D$ (or of hermitian $\gamma_5 D$),

(2) count the real eigenvalues of $D_{\rm w}$ in the region
$0<=\lambda(D_{\rm w})<\mu$, with chirality signs,

(3) trace the ``level crossings'' of 
$H(\mu')=\gamma_5(\mu'-D_{\rm w})$ while $\mu'$ runs from $0$
to $\mu$.

Note that although we rely here on the exact index theorem on the
lattice, the value of the topological charge determined through
$D$ with $A=\mu-D_{\rm w}$ is the same as the method using
$D_{\rm w}$ suggested a long time ago \cite{SV,Itoh}.
The last method has also been used in Ref.~\cite{OLF} 
in the framework of the overlap formalism.

This definition of the topological charge works well for small
instantons, as expected. But how does it work for typical MC
configurations? Does $D$ (generated using $D_{\rm w}$) give a `better'
definition than the geometrical one? Hernandez \cite{Hern} argues
that the answer is no. Gattringer and Hipp \cite{GattH} simulated
SU(2) on a $12^4$ lattice at $\beta=2.4$ and found that the value of 
the topological charge, $\nu$ on individual configurations depends 
strongly on the parameter $\mu$ since there are usually many 
real eigenvalues in the region of interest, and only very few
configurations have a gap in the spectrum where $\nu$ does
not depend on $\mu$ chosen in a reasonable range.
They also note that a naive inclusion of the clover term
makes the situation worse. The same conclusion is found 
by DeGrand, A.~Hasenfratz and Kov\'acs \cite{DHK}.
This last fact is not too surprising since the corresponding
artifact is not an O($a$) effect.

Note, however, that in the average this definition of charge works
better: Edwards, Heller and Narayanan \cite{Edw} found that the
topological susceptibility stays roughly constant (independent
of $\mu$) for $\mu\ge 1.0$. Here I would like to point out that 
the $\mu$-independence of $\langle \nu^2 \rangle$ does not
necessarily mean the `true' value: if 
$\nu=\nu_{\rm true}+\,{\rm noise}$ then 
$\langle \nu^2 \rangle > \langle \nu_{\rm true}^2 \rangle $.
A good definition of the topological charge (at a given lattice
spacing) should give relatively unambiguous definition of $\nu$
on individual configurations. In other words, the value of the
lattice spacing $a$ when a GW Dirac operator gives a `good'
definition of the topological charge could strongly depend on
the choice of $D$. With the presently used lattice spacings
the original choice of Neuberger (using $D_{\rm w}$)
seems to produce a topological charge which is too ambiguous.

On the other hand, measuring the topological charge with the
FP action in the Schwinger model \cite{Schw}
gives quite convincing results. For 4d gauge theories, however,
the situation is less satisfactory \cite{DHZK} -- mostly
because of technical problems to perform the required multigrid
minimization on the gauge configurations.

For QCD a good strategy would be a compromise: take a Dirac operator
$D_0$ with good properties (small cut-off effects, nearly satisfying
the GW relation,\ldots) and build $D$ from it, which in turn could be
used in the definition of the topological charge. It would have
smaller O($a^2$) artifacts, better locality, fewer `dangerous' 
configurations, larger gap in the real eigenvalues hence less 
ambiguity of $Q_{\rm top}$, but will also need more programming 
effort and extra computational overhead.

Measuring the topological charge density using
$q(x)=\frac{1}{2}{\rm tr}(\gamma_5 D(x,x;U))$ is perhaps too 
time consuming to be feasible.

\section{On chiral gauge theories}

The question whether it is possible to define chiral gauge theories 
(like the Standard Model) on the lattice is a purely theoretical
problem, but of great importance. Much has been done in this direction
and has been reported earlier \cite{ChGTconf,Neub98}.
Nevertheless, there remain several open questions.

A GW Dirac operator seems to be a good starting point: it has exact
gauge invariance and one can define Weyl fermions 
(cf. eq.~(\ref{50},\ref{51})).
Here I shall report briefly about new developments 
(partially yet unpublished) by L\"uscher \cite{L-ChGT}
based on the properties of GW Dirac operators.
The discussion will touch only a few interesting points 
which could be easily understood on the basis of previous sections.

Consider fermion fields which are purely left handed 
(see eq.~(\ref{50}))
\begin{equation}
\label{76}
\hat{P}_- \psi=\psi\,,~~ \overline{\psi}P_+ = \overline{\psi} \,.
\end{equation}
These are local, gauge invariant conditions. Introduce a basis
for $\psi(x)$ and $\overline{\psi}(x)$ through relations
\begin{eqnarray}
& & \hat{P}_- v_j=v_j\,,~~ \psi(x)=\sum_j v_j(x) c_j \,, \label{77} \\
& & \overline{v}_k P_+ =\overline{v}_k\,,~~ 
\overline{\psi}(x)=\sum_k \overline{c}_k \overline{v}_k(x) \,,
\label{78}  
\end{eqnarray}
where $c_j$ and $\overline{c}_k$ are Grassmann variables.
In this basis the fermion action reads
\begin{equation}
\label{79}
S_{\rm F}=\overline{\psi}D\psi = 
\sum_{kj} \overline{c}_k M_{kj} c_j
\end{equation}
with $M_{kj}=\sum_x \overline{v}_k D v_j$.

The fermion propagator (when there are no zero modes of $D$)
is given by
\begin{equation}
\label{80}
\left\langle \psi(x)\overline{\psi}(y)\right\rangle_{\rm F} =
\langle 1 \rangle_{\rm F} \left( \hat{P}_- D^{-1} P_+\right)_{xy} \,.
\end{equation}
It is easy to show that this is chiral in the generalized sense.
Note that 
\begin{equation}
\label{81}
 \hat{P}_- D^{-1} P_+ =  P_- D^{-1} P_+ + \frac{1}{2}P_+ \,.
\end{equation}
The first term on the rhs. is chiral in the usual sense 
(with $\gamma_5$) while $P_+$ is a contact term.

Consider a chiral gauge theory with fermion multiplet 
$\alpha=1,2,\ldots,N$, with integer charges $e_\alpha$.
For the number of zero modes $(n_+)_\alpha$, $(n_-)_\alpha$
the index theorem implies
\begin{equation}
\label{82}
(n_-)_\alpha - (n_+)_\alpha = e_\alpha^2 \nu
\end{equation}
since $F\tilde{F}$ is replaced by $e_\alpha^2 F\tilde{F}$ for smooth
gauge fields when $e_\alpha \ne 1$.

A simple but important observation is that the number of $\psi$
and $\overline{\psi}$ fields are not necessarily equal.
They are given by ${\rm Tr}(\hat{P}_-)$ and ${\rm Tr}(P_+)$,
respectively. For their difference we have
\begin{equation}
\label{83}
{\rm Tr}\left(\hat{P}_-\right) - {\rm Tr}\left(P_+\right)
= \frac{1}{2}{\rm Tr}\left( \gamma_5 D\right)
= \nu \sum_\alpha e_\alpha^2 \,.
\end{equation}
To illustrate this, consider a background field with $\nu=1$, 
where $n_-=1$, $n_+=0$ and $n_-'=0$, $n_+'=1$.
Then for the $\lambda=0$ mode
\begin{equation}
\label{84}
\gamma_5 v = \hat{\gamma}_5 v = -v
\end{equation}
while for the eigenvector $v'$ at $\lambda=2$
\begin{equation}
\label{85}
\gamma_5 v' = -\hat{\gamma}_5 v' = v'
\end{equation}
According to eqs.~(\ref{76}-\ref{78}), at $\lambda=0$ we have
only $\psi$ (but not $\overline{\psi}$) while at $\lambda=2$
both $\psi$ and $\overline{\psi}$ are present since $\gamma_5$
and $\hat{\gamma}_5$ differ by sign at this point.
Obviously for complex values of $\lambda$ 
both $\psi$ and $\overline{\psi}$ are present again.
Altogether there is one more $\psi$ than $\overline{\psi}$.
Of course, this statement depends on the topological charge of 
the gauge field. 

As a consequence the fermionic average 
$\left\langle{\cal O}\right\rangle_{\rm F}$
of any operator in a given background gauge field vanishes
unless the fermionic number of type $\alpha$
(the number of $\psi_\alpha$'s minus the number of 
$\overline{\psi}_\alpha$'s) in ${\cal O}$ equals to
$(n_-)_\alpha - (n_+)_\alpha=\nu\sum_\alpha e_\alpha^2$.

Perform now (for the U(1) case) a global gauge transformation
$g(x)=\exp(i\omega)$. This modifies ${\cal O}$ by a phase factor
$\exp(i\phi)$ where 
$\phi=\omega \sum_\alpha e_\alpha (n_- - n_+)_\alpha
=\omega \sum_\alpha e_\alpha^3$.
 From the other side a global U(1) gauge transformation does not
change the gauge fields at all, hence invariance under this
transformation requires
\begin{equation}
\label{86a}
\sum_\alpha e_\alpha^3 = 0 \,,
\end{equation}
i.e. the condition of anomaly cancellation of the continuum theory.

These observations, however, constitute only the starting point
of a difficult mathematical problem -- to find a satisfactory
choice of the fermionic integration measure
(``chiral determinant''), i.e. of the basis $v_j(x;U)$.

L\"uscher has proven that it is possible to construct
a basis for the U(1) case that satisfies the following requirements:

(1) The fermionic expectation values 
 $\left\langle {\cal O} \right\rangle_{\rm F}$ are smooth functions of $U$

(2) The field equations (obtained by some change of variables)
 are {\em local}:
\begin{equation}
\label{87}
\left\langle \delta {\cal L}(y) {\cal O}(x_1){\cal O}(x_2)\ldots
\right\rangle = 0
\end{equation}
if $y$ is far from $x_1,x_2,\ldots$.

(3)
$\left\langle {\cal O} \right\rangle_{\rm F}$ is gauge invariant 
if ${\cal O}$ is.

Note that the last property may not be necessary to require.
In principle, it is possible that after the leading violation
of gauge invariance is cancelled by the condition
$\sum_\alpha e_\alpha^3=0$, the rest is small and goes away in the
continuum limit, without the need of extra counter terms.
This possibility is based on a simple but surprising result 
of Foerster, Nielsen and Ninomiya \cite{FNN}, who show that 
gauge invariance on the lattice is robust in the sense that the 
effect of a sufficiently small violating term vanishes in the 
continuum limit. The constructions of the overlap and
gauge fixing \cite{GF} formalisms assume this scenario.
It is clear, however, that it is much safer -- if possible --
to preserve chiral gauge invariance at finite lattice spacing.

\section{Summary}

It has been illustrated that Dirac operators satisfying the GW
relation provide an elegant solution to the chirality problem
in lattice QCD. In particular, there is no additive quark mass
renormalization, and Ward identities, soft pion theorem, etc. are valid,
there are no O($a$) artifacts.

It is pointed out that there is a large freedom in choosing 
a GW Dirac operator, and the choice should be optimized 
for locality, smallness of artifacts, etc.

For practical implementations of these ideas new algorithms are
needed. That the task is feasible is illustrated by MC
calculations in the domain wall approach \cite{Blum98}
which could be considered as a particular approximation
to a GW Dirac operator. Not restricting itself to the given
ideology could open ways to more efficient algorithms.

Finally, as shown by L\"uscher's construction for the U(1)
case, it seems to be possible to define chiral gauge theories on the
lattice based on a GW Dirac operator.

\noindent {\bf Acknowledgements}. I would like to thank Sinya Aoki,
Peter Hasenfratz, Victor Laliena, Martin L\"uscher and Peter Weisz 
for illuminating discussions.

\end{document}